\documentclass[aps,prd,nofootinbib,groupedaddress,twocolumn,floatfix]{revtex4}
\usepackage{graphicx}
\usepackage{epsfig}
\usepackage{dcolumn}
\usepackage{amsmath}
\def\Journal#1#2#3#4{{#1} {\bf#2}, #3 (#4)}

\def\NPA{{\rm Nucl. Phys.} A}
\def\NPB{{\rm Nucl. Phys.} B}
\def\PLB{{\rm Phys. Lett.}  B}
\def\PRL{\rm Phys. Rev. Lett.}
\def\PRD{{\rm Phys. Rev.} D}
\def\PRC{{\rm Phys. Rev.} C}

\def\EPJC{{\rm Eur. Phys. J.}C}

\def\la{\langle}
\def\ra{\rangle}

\def\be{\begin{equation}}
\def\ee{\end{equation}}
\def\bea{\begin{eqnarray}}
\def\eea{\end{eqnarray}}
\begin{document}
\title{Conformal Symmetry and Pion Form Factor: Space- and Time-like
Region}
\author{ Ho-Meoyng Choi$^{a}$ and Chueng-Ryong Ji$^{b}$\\
$^a$ Department of Physics, Teachers College, Kyungpook National University,
     Daegu, Korea 702-701\\
$^b$ Department of Physics, North Carolina State University,
Raleigh, NC 27695-8202}
\begin{abstract}
We extend a recent analysis of the pion electromagnetic form factor
constrained by the conformal symmetry to explore the time-like region.
We show explicitly that the time-like form factor obtained by the analytic
continuation of the space-like form factor correctly satisfies the dispersion
relation. Our results indicate that the quark spin and dynamical mass effects
are crucial to yield the realistic features of the vector meson dominance
phenomena.
\end{abstract}


\maketitle

\section{Introduction}
One of the most significant theoretical advances in recent years has been
the application of AdS/CFT correspondence~\cite{Mal99} between
string theories defined on the 5-dimensional Anti-de Sitter(AdS) space-time
and conformal field theories(CFT) in physical space-time.
Although QCD is not itself a conformal theory, it may possess
an approximate conformal symmetry in the domain where the QCD coupling
is approximately constant and quark masses can be neglected. Based on the
premise that QCD belongs to this class, theoretical development
of AdS/CFT to QCD, often referred as ``AdS/QCD" or
``holographic QCD", has been very popular in recent
years. The resulting AdS/QCD model gives predictions for hadron
spectroscopy~\cite{Mass} and a description of the quark structure of
hadrons~\cite{PS02,BT04} which has scale invariance and dimensional
counting~\cite{BL} at short distances, together with color confinement
at large distances.

The AdS/QCD correspondence is particularly relevant for the description of
hadronic form factors and also provides a convenient framework for analytically
continuing the space-like results to the time-like region.
The form factors have been studied within the holographic
approach~\cite{PS02}, and the connection between the AdS/QCD
approach~\cite{PS02,BT04} and the
usual light-front formalism for hadronic form factors was proposed
in~\cite{BT06} and discussed in~\cite{Rad06}. Some other recent applications
of different AdS/QCD models(hard-wall or soft-wall models)to the
form factors of hadrons can be found in~\cite{GuyStan,BT07,BT_Pi,GR07,KL}.
In AdS/QCD one breaks the conformal symmetry by impeding the ability of the
field $\Psi(x,z)$ to penetrate deeply into the bulk, which leads to an
explanation of confinement and more generally constrains the model's
infrared(IR) behavior. This may be accomplished by imposing a hard cutoff
and appropriate boundary conditions on $\Psi(x,z)$ at some finite value
of fifth dimension $z=z_0$~\cite{PS02}(hard-wall model) or by imposing
a soft IR cutoff with an oscillator-type potential in the action for
large $z$~\cite{KK}(soft-wall model).

The connection between the AdS/QCD and the light-front approaches allows
one to compute the analytic form of frame-independent light-front
wavefunctions $\psi_{n/h}$ of hadrons in physical space-time~\cite{BT06},
thus providing a relativistic description of hadrons in QCD at the amplitude
level. The pion electromagnetic form factor has been exemplified by AdS/QCD,
in particular, the power-law behavior of the form factor
$F_\pi(Q^2)\sim 1/Q^2$ is well reproduced by the soft-wall AdS/QCD
model(or Gaussian model)~\cite{GuyStan}.
The key ingredient in this correspondence is the conformal symmetry valid
in the negligible quark mass. Using the connection between the AdS/QCD
and the light-front approaches in the calculation of the hadronic form
factors~\cite{GuyStan}, we calculated~\cite{CJ06} the pion form factor in
our light-front quark model(LFQM) taking into account a momentum-dependent
dynamical quark mass. We confirmed that the power-law behavior of the pion
form factor is indeed attained by taking into account a momentum-dependent
dynamical quark mass which becomes negligible at large momentum region.
Our result~\cite{CJ06} was consistent with an
important point of AdS/QCD prediction, namely, the holographic wavefunction
contains the contribution from all scales up to the confining scale.
We also have shown that the broader shape of the pion distribution amplitude
increases the magnitude of the leading twist perturbative QCD(PQCD)
predictions for the pion form factor by a factor of 16/9 compared to the
prediction based on the asymptotic form.
Very recently, Brodsky and de Teramond~\cite{BT_Pi} extended their AdS/QCD
model to obtain the time-like pion form factor by doing analytic continuation
of the space-like result to the time-like region. However, the
dispersion relation of the time-like form factor $F(q^2)={\rm Re}F(q^2)+ i{\rm Im}F(q^2)$
has not yet been analyzed explicitly.
We thus show in this work that the time-like form factor
obtained by the analytic continuation of the space-like form factor correctly
satisfies the dispersion relation.
Working in the framework of the LFQM that takes into
account a momentum-dependent dynamical quark mass~\cite{CJ06}, we
further extend our previous analysis of the space-like pion form factor to the
time-like region and compare with the result obtained from the AdS/QCD
model~\cite{BT_Pi}. Our comparative analysis reveals the effects from
the quark spin and dynamical mass.

The paper is organized as follows. In Sec. II, we present the soft
contribution to the pion form factor using the LFQM. In Sec. III, we present
the space- and time-like pion form factor in the conformal symmetry limit. We
also compare our LFQM results with the soft-wall AdS/QCD model
calculation~\cite{BT_Pi} to investigate the quark spin structure inside the
pion. In Sec. IV, we show our numerical results of both space- and
time-like pion form factor. The direct calculation of the pion form factor
is shown to be in excellent agreement with the result obtained from the
dispersion relation. Summary and conclusion follow in Sec. V.

\section{Pion Form Factor in LFQM}

The elastic pion form factor is related to pion electromagnetic current
by the following equation:
\bea\label{pion_ma}
\la P'|J^\mu(0)|P\ra = (P' + P)^\mu F_\pi (Q^2).
\eea
As usual, our calculation will be carried out using the Drell-Yan-West
frame($q^+$=$q^0+q^3$=0) where
$q^2=(P-P')^2=q^+q^- -{\bf q}^2_\perp =-Q^2$, i.e. $Q^2>0$ is the
space-like momentum transfer. In this frame, the matrix element of the current
can be expressed as convolution integral in terms of the light-front
wavefunction and the pion form factor can be written for the
``+"-component of the current $J^\mu$ as follows(see~\cite{CJ99} for
more detailed calculation):
\bea\label{soft_ff}
F_{\pi}(Q^2)\hspace{-0.2cm}&=&\hspace{-0.2cm}
\int^1_0 \hspace{-0.2cm}dx\int d^2{\bf k}_\perp
\sqrt{\frac{\partial k_z}{\partial x}}\phi_R(x,{\bf k}_\perp)
\sqrt{\frac{\partial k'_z}{\partial x}}\phi_R(x,{\bf k'}_\perp)
\nonumber\\
&&\times
\frac{m^2 + {\bf k}_\perp\cdot{\bf k'}_\perp}
{\sqrt{m^2+{\bf k}^2_\perp}\sqrt{m^2+{\bf k'}^2_\perp}},
\eea
where ${\bf k'}_\perp = {\bf k}_\perp + (1-x){\bf q}_\perp$ and the
factors $m^2$ and ${\bf k}_\perp\cdot{\bf k'}_\perp$ in the numerator come
from the ordinary helicity($\lambda+\bar\lambda=0$) and the higher
helicity($\lambda+\bar\lambda=\pm 1$) components of the spin-orbit wave
function ${\cal R}_{\lambda\bar\lambda}(x,{\bf k}_\perp)$, respectively.
Here, the explicit form of ${\cal R}_{\lambda\bar\lambda}(x,{\bf k}_\perp)$
is obtained by the interaction-independent
Melosh transformation~\cite{Mel}
from the ordinary equal-time static spin-orbit wave function assigned by
the quantum numbers $J^{PC}=0^{-+}$.
The radial wave function is given by
\bea\label{radial}
\phi_R(x,{\bf k}_\perp)= \sqrt{\frac{1}{\pi^{3/2}\beta^3}}
\exp(-{\vec k}^2/2\beta^2),
\eea
where the gaussian parameter $\beta$ is related with the size of pion and
the three momentum squared ${\vec k}^2$ can be represented by
the light-front (LF)
variables, i.e.
\bea\label{ksquare}
{\vec k}^2 = \frac{{\bf k}_\perp^2 + m^2}{4x(1-x)}-m^2,
\eea
for the quark mass $m_u = m_d = m$.
For additional factor $\sqrt{\partial k_z/\partial x}$
to the radial wavefunction $\phi_R(x,{\bf k}_\perp)$
(similarly $\sqrt{\partial k'_z/\partial x}$ to
$\phi_R(x,{\bf k'}_\perp$))
is the Jacobian of the variable transformation
$\{x,{\bf k}_\perp\}\to {\vec k}=({\bf k}_\perp, k_z)$
due to the normalization of the radial wavefunction~\cite{CJ_Jacob}.
For the low momentum transfer phenomenology in LFQM, it is customary
to take a constant constituent quark mass $m$ as a mean value of
the momentum dependent dynamical quark mass at low momentum region.
The momentum dependence of the dynamical quark mass in the spacelike
momentum region has been discussed in lattice QCD~\cite{Wil} as well as
in other approaches such as Dyson-Schwinger~\cite{DS1,DS2} and
instanton~\cite{Dor} models.
Also, the difference between spacelike and timelike meson form factors
at large momentum transfer was discussed in the framework of PQCD with
Sudakov effects included~\cite{GP}.
Matching between the low momentum LFQM prediction and the large momentum
PQCD prediction is a highly nontrivial task which goes beyond the scope of
our present work. Nevertheless, as we discussed in~\cite{CJ06},
one should understand $m$ as a function of
$Q^2$ in principle  although in practice $m(Q^2)$ for the low $Q^2$
phenomenology can be taken as a constant constituent quark mass.
In our previous analysis~\cite{CJ06}, we took the simple parametrization
of the quark mass evolution $m(Q^2)$ in space-like momentum region as
$m(Q^2)=m_0 + (m_c - m_0)(1+ e^{-0.2})/[1+ e^{(Q^2-0.2)}]$, where
$m_0=5$ MeV and $m_c=220$ MeV represent the current(at high $Q^2$) and
constituent(at low $Q^2$) quark mass, respectively.
In this work, we take the same phenomenological form of $m(Q^2)$
used in~\cite{CJ06}.

We note~\cite{CJ06} that
$\phi_R(x,{\bf k}_\perp) \phi_R(x,{\bf k'}_\perp)$ in
Eq.~(\ref{soft_ff}) provides a mass-dependent weighting factor
$e^{-\frac{m^2}{4x(1-x)\beta^2}}$
which severely suppresses the contribution from the endpoint region of
$x \rightarrow 0$ and 1 unless $m \rightarrow 0$.
This weighting factor leads to the gaussian fall-off of $F_\pi(Q^2)$
at high $Q^2$ region for the constant constituent quark mass which breaks
the conformal symmetry.
When the conformal symmetry limit ($m \rightarrow 0$) is taken, however,
there is no such suppression of the endpoint region and the high $Q^2$
behavior of the form factor dramatically changes from a gaussian fall-off to
a power-law reduction. In the next section, we shall elaborate the
pion form factor to explain why the power-law behavior
attained in our Eq.~(\ref{soft_ff}) is not accidental but a consequence of
the constraint taken from the conformal symmetry.

\section{Space- and Time-like Pion Form Factor in Conformal Symmetry}
In order to facilitate our calculation, we change the momentum
variables as ${\bf l}_\perp = {\bf k}_\perp + (1-x){\bf q}_\perp/2$
and $\xi^2=(1-x)^2{\bf q}^2_\perp/4$. Then, the pion form
factor(Eq.~(\ref{soft_ff})) in the conformal symmetry limit($m\to 0$)
is obtained as
\bea\label{FF2}
F_\pi(Q^2)&=& \frac{1}{2\pi^{3/2}\beta^3}
\int^1_0 dx\int^\infty_0 d{\bf l}^2_\perp\int^{2\pi}_0 d\phi
{\cal J}{\cal M}
\nonumber\\
&&\times
{\rm exp}\biggl[-\frac{ {\bf l}^2_\perp +\xi^2}{4\beta^2x(1-x)}\biggr],
\eea
where $\phi$ is the cosine angle between ${\bf l}_\perp$ and
${\bf q}_\perp$. The Melosh factor ${\cal M}$ coming from the spin structure
of the pion and the product of two Jacobi factors ${\cal J}$ are given by
\bea\label{Mel}
{\cal M}&=& \frac{{\bf k}_\perp\cdot{\bf k'}_\perp}
{\sqrt{{\bf k}^2_\perp}\sqrt{{\bf k'}^2_\perp}}
= \frac{ {\bf l}^2_\perp-\xi^2 }
{\sqrt{ ({\bf l}^2_\perp +\xi^2)^2 - 4{\bf l}^2_\perp\xi^2{\rm cos}^2\phi}},
\eea
and
\bea\label{Jacobi}
{\cal J}
&=& \sqrt{\frac{\partial k_z}{\partial x}}
\sqrt{\frac{\partial k'_z}{\partial x}}
=\frac{ [({\bf l}^2_\perp +\xi^2)^2
- 4{\bf l}^2_\perp\xi^2{\rm cos}^2\phi]^{1/4}}{4[x(1-x)]^{3/2}},
\nonumber\\
\eea
respectively. The exponential function in Eq.~(\ref{FF2}) comes from the
product of two radial wave functions $\phi_R(x,{\bf k}_\perp)$ and
$\phi_R(x,{\bf k'}_\perp)$.

For the quark spin $s=1/2$ case, all of the factors
${\cal J}, {\cal M}$ and $\phi_R(x,{\bf k}_\perp)\phi_R(x,{\bf k'}_\perp)$
should be kept in Eq.~(\ref{FF2}) and the pseudoscalar pion form factor
respecting conformal symmetry($m=0$)
behaves as $F^{s=1/2}_\pi(Q^2)\propto 1/Q^4$ at large $Q^2$.
For the scalar quark ($s=0$) case, however, the Melosh factor is turned
off (i.e. ${\cal M}=1$) and the corresponing pion form factor in the
conformal limit ($m=0$) behaves as
$F^{s=0}_\pi(Q^2)\propto 1/Q^2$.
The latter case ($F^{s=0}_\pi(Q^2)\propto 1/Q^2$) is shown~\cite{CJ06}
to be equivalent to the soft-wall
AdS/QCD result at large momentum transfer~\cite{BT07,BT_Pi}.

Let us now explore more of the spin and mass evolution effects
to the pion form factor in space- and time-like regions.
In the conformal limit($m=0$) of the scalar quark
(i.e. ${\cal M}=1$ in Eq.~(\ref{FF2})) case,
the form factor $F^{s=0}_\pi(Q^2)$ after the $\phi$-integration has the following form
\bea\label{Form_WOM}
F^{s=0}_\pi(Q^2)&=& \frac{1}{4\beta^3\sqrt{\pi}}
\int^1_0 \frac{dx}{[x(1-x)]^{3/2}}\int^\infty_0 d{\bf l}^2_\perp
e^{-\frac{ {\bf l}^2_\perp +\xi^2}{4\beta^2x(1-x)}}
\nonumber\\
&&\times
\sqrt{{\bf l}^2_\perp +\xi^2}\;
{_2F_1}\biggl(-\frac{1}{4},\frac{1}{2},1;
\frac{4{\bf l}^2_\perp\xi^2}{({\bf l}^2_\perp+\xi^2)^2}\biggr),
\eea
where ${_2F_1}(a,b,c;x)$ is the hypergeometric function with the range of
convergence $|x|<1$ and $x=1$, for $c> a+b$, and $x=-1$, for $c>a+b-1$.
Changing the variables in Eq.~(\ref{Form_WOM}) as
${\bf l}^2_\perp=\xi^2\tan^2\theta$ and
$4{\bf l}^2_\perp\xi^2/({\bf l}^2_\perp+\xi^2)^2=\sin^2 2\theta$,
we obtain
\bea\label{Form_WOM2}
F^{s=0}_\pi(Q^2)&=&
\frac{(Q^2)^{3/2}}{16\beta^3\sqrt{\pi}}
\int^1_0 dx\biggl[\frac{1-x}{x}\biggr]^{3/2}\int^{\pi/2}_0 d\theta
g(\theta)
\nonumber\\
&&\times
e^{-\frac{(1-x)Q^2}{16\beta^2x\cos^2\theta}},
\eea
where $g(\theta)=(\sin\theta/\cos^4\theta){_2F_1}(-1/4,1/2,1; \sin^22\theta)$.

After the $x$-integration, we could further
reduce Eq.~(\ref{Form_WOM2}) to the following 1-dimensional form
\bea\label{In_t}
F^{s=0}_\pi(Q^2)\hspace{-0.2cm}&=& \hspace{-0.2cm}
\frac{(Q^2)^{3/2}}{16\beta^3}
\int^{\pi/2}_0 d\theta g(\theta)
\nonumber\\
&&\times
\biggl\{\frac{(1+z)}{\sqrt{z}}
-\frac{1}{2}(3+2z)e^z\Gamma(\frac{1}{2},z)
\biggr\},
\eea
where $z=Q^2/16\beta^2{\rm cos}^2\theta$ and
$\Gamma(\frac{1}{2},z)=\sqrt{\pi}{\rm Erfc}(z)$. Here the incomplete gamma function,
$\Gamma(\frac{1}{2},z)$, has a branch cut discontinuity in the complex $z$ plane
running from $-\infty$ to 0. This is the reason why the pion form factor
is complex in time-like
region($q^2=-Q^2$) but real in space-like region. We now obtain the time-like
form factor $F_\pi(q^2)$ by changing $Q^2$ to $-q^2$ in the form factor given
by Eq.~(\ref{In_t}), where the imaginary part is obtained as
\be\label{Fpi_im}
{\rm Im}[F^{s=0}_\pi(q^2)]=
\frac{(q^2)^{3/2}\sqrt{\pi}}{32\beta^3}
\int^{\pi/2}_0 d\theta g(\theta)
(3-2z')e^{-z'},
\ee
where $z'=q^2/16\beta^2{\rm cos}^2\theta$.

We find numerically that Eq.~(\ref{Form_WOM}) is equivalent to
the soft-wall AdS/QCD result in the large $Q^2$ limit where
the unconfined bulk-to-boundary propagator can be used~\cite{BT_Pi},
i.e. $F^{\rm AdS/QCD}_{(u.c.)}(Q^2)=\int^1_0 dx {\rm exp}[-(1-x)Q^2/4\kappa^2x]$
(see Eq.(E.8) in~\cite{BT_Pi}). This leads to the following analytic form
in the space-like region
\bea\label{AdS1}
F^{\rm AdS/QCD}_{(u.c.)}(Q^2) &\simeq&
\frac{1 - e^{\frac{Q^2}{4\kappa^2}}Q^2\Gamma(0,\frac{Q^2}{4\kappa^2})}
{4\kappa^2},
\eea
where the subscript (u.c.) denotes the result of the unconfined
current decoupled from the dilaton field.
The imaginary part of the time-like $F^{\rm
AdS/QCD}_{(u.c.)}(q^2)$
is given by
\bea\label{Im_F}
{\rm Im}[F^{\rm AdS/QCD}_{(u.c.)}(q^2)]&=&-\pi
q^2\frac{e^{-\frac{q^2}{4\kappa^2}}}{4\kappa^2}.
\eea
In the time-like region, the following dispersion relation should be satisfied
\bea\label{Re_F}
{\rm Re}[F_\pi(q^2_0)]&=& \frac{1}{\pi}P
\int^\infty_0\frac{{\rm Im}[F_\pi(q^2)]}{q^2-q^2_0}dq^2,
\eea
where $P$ denotes the Cauchy principal value.

In our numerical calculation, we find that the equivalence between our
$F_\pi^{s=0}(Q^2)$ in conformal($m=0$) limit and
$F^{\rm AdS/QCD}_{(u.c.)}(Q^2)$ is
achieved by matching our $\beta=0.173$ GeV
with $\kappa=0.4$ GeV in~\cite{BT_Pi}(i.e. $\kappa\simeq 2.3\beta$).
Comparing the gaussian dependence
of our model wave function $\Psi(x,{\bf k}_\perp)\sim
\sqrt{\partial k_z/\partial x}e^{-{\bf k}^2_\perp/2(2\beta)^2x(1-x)}$
in $m=0$ limit and
the unconfined version of the
soft-wall AdS/QCD model wave function
$\Psi(x,{\bf k}_\perp)\sim
e^{-{\bf k}^2_\perp/2\kappa^2x(1-x)}/\sqrt{x(1-x)}$~\cite{BT_Pi},
we understand the relation between $\beta$ and $\kappa$ which is not
exactly $\kappa=2\beta$ but $\kappa\simeq 2.3\beta$ due to the different
free factors. Other than such fine details, they are essentially equivalent
to each other.
The pion form factor in Eq.~(\ref{AdS1}) respecting the conformal symmetry
shows the power-law behavior of $F_\pi(Q^2)\to 4\kappa^2/Q^2$
at large $Q^2(\gg \kappa^2)$.
The equivalence between our result
without the Melosh factor in Eq.~(\ref{Form_WOM})
and that of the soft-wall AdS/QCD model in the large $Q^2$ limit given by
Eq.~(\ref{AdS1}) is assured when our LFQM respects the conformal symmetry.

We note that the authors in~\cite{BT_Pi} also
derived the pion form factor in the presence of a dilaton field
in AdS space and presented the analytic solution of the modified
wave equation for the confined bulk-to-boundary propagator.
This corresponds to the well-known vector dominance model(VDM)
with the leading $\rho$ resonance, i.e.
\be\label{AdS_c}
F^{\rm AdS/QCD}_{(confined)}(Q^2)=\frac{4\kappa^2}{4\kappa^2 + Q^2}.
\ee
In the timelike region, one may modify Eq.(\ref{AdS_c}) to
introduce a finite width (e.g. $\Gamma = 100$ MeV in~\cite{BT_Pi})
to compare with the pion form factor data near the $\rho$ peak.

Although the analytic solution for the soft-wall unconfined current
(Eq.~(\ref{AdS1})) is not much different from that for the soft-wall confined
current(Eq.~(\ref{AdS_c})) at large momentum transfer, the latter solution fits the
low $Q^2$ region much better than the former one. This implies that the dilaton
background only affects the low $Q^2$ region but shows the same $1/Q^2$ power-law
behavior of the pion form factor as that for the unconfined current.
In our LFQM analysis, we consider both spin and mass evolution effects from the
constituent quark and discuss the corresponding power-law behaviors in the
pion form factor.

In order to see the spin effect on the pion form factor,
we should include the Melosh factor ${\cal M}$ given by Eq.~(\ref{Mel}).
In this pseudoscalar($s=1/2$) case, the pion form factor $F^{s=1/2}_\pi(Q^2)$
given by Eq.~(\ref{FF2}) has the following form
\bea\label{Form_WM}
F^{s=\frac{1}{2}}_\pi(Q^2)\hspace{-0.2cm}&=&\hspace{-0.2cm}
\frac{1}{4\beta^3\sqrt{\pi}}
\int^1_0 \frac{dx}{[x(1-x)]^{3/2}}\int^\infty_0 d{\bf l}^2_\perp
e^{-\frac{ {\bf l}^2_\perp +\xi^2}{4\beta^2x(1-x)}}
\nonumber\\
&&\times
\frac{{\bf l}^2_\perp -\xi^2}{\sqrt{{\bf l}^2_\perp +\xi^2}}
{_2F_1}\biggl(\frac{1}{4},\frac{1}{2},1;
\frac{4{\bf l}^2_\perp\xi^2}{({\bf l}^2_\perp+\xi^2)^2}\biggr),
\eea
after the $\phi$-integration.
Following the same procedure as in the case of scalar pion case, we obtain
the pseudoscalar pion form factor $F^{s=1/2}_\pi(Q^2)$ in conformal
limit by changing $g(\theta)$ in Eq.~(\ref{In_t})(and
Eq.~(\ref{Fpi_im})) to
$g'(\theta)=-(\sin\theta\cos 2\theta/\cos^4\theta)
{_2F_1}(1/4,1/2,1; \sin^22\theta)$.
We should note that the $Q^2$ behavior of $F^{s=1/2}_\pi(Q^2)$
is quite different from that of $F^{s=0}_\pi(Q^2)$
since the $\theta$ variable is related with the light-front momentum variables
$(x,{\bf k}_\perp)$ as well as the momentum transfer $Q^2$.
As discussed earlier, the
pseudoscalar pion form factor behaves $F^{s=1/2}_\pi(Q^2)\propto 1/Q^4$ at
large $Q^2$ compared to the scalar pion case,
$F^{s=0}_\pi(Q^2)=F^{\rm AdS/QCD}_{(u.c.)}(Q^2)\propto 1/Q^2$.
\begin{figure}
\vspace{0.8cm}
\includegraphics[height=1.5in,width=1.5in]{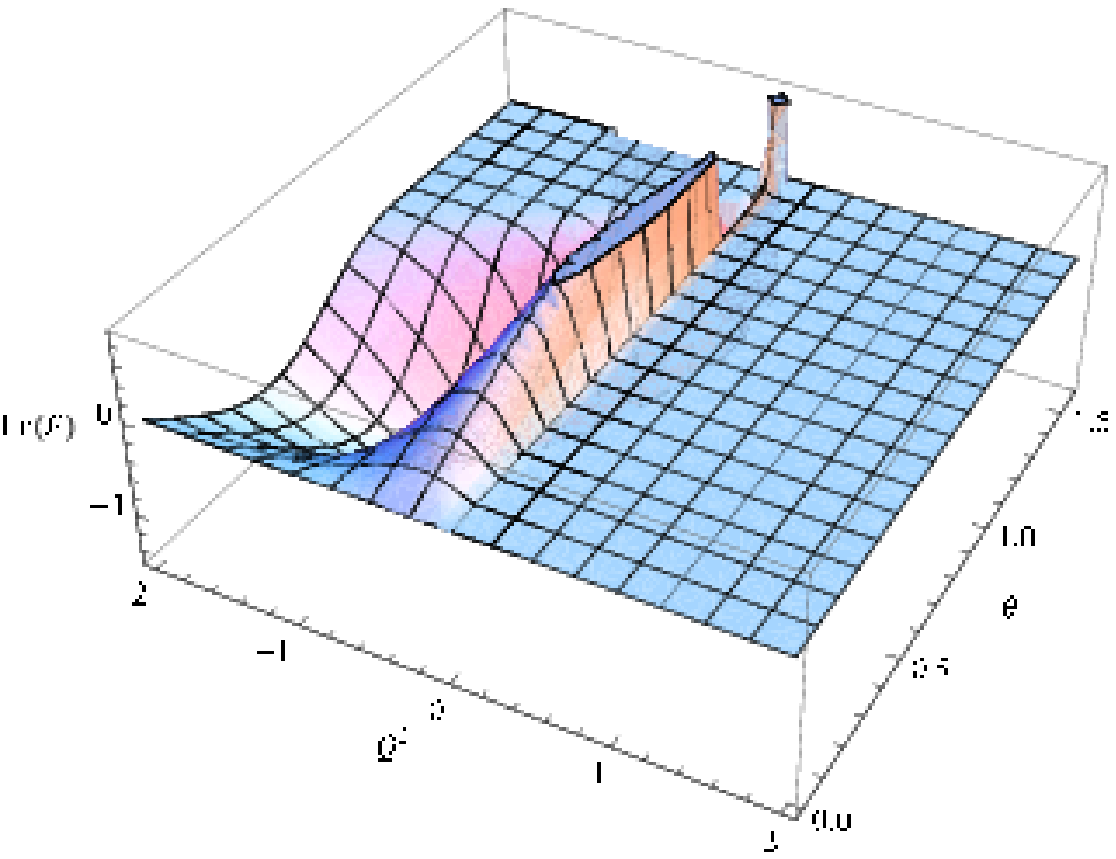}
\hspace{0.5cm}
\includegraphics[height=1.5in,width=1.5in]{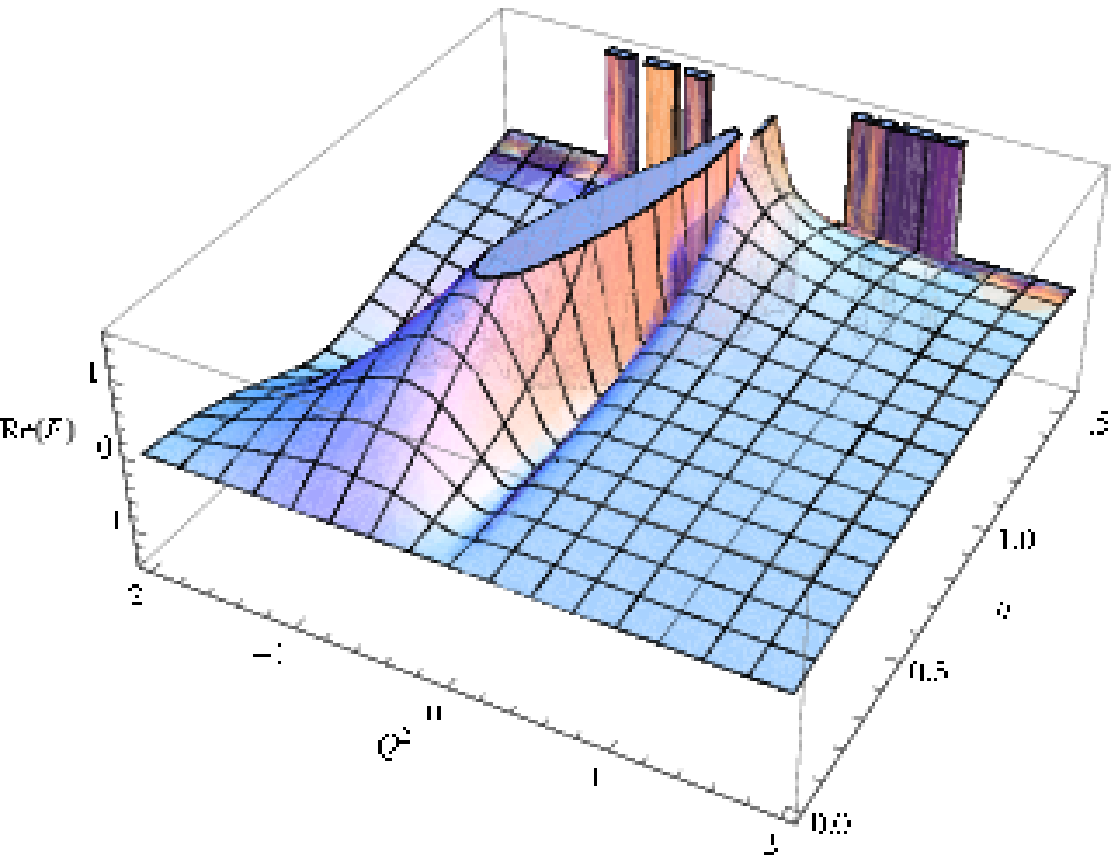}
\\
\includegraphics[height=1.5in,width=1.5in]{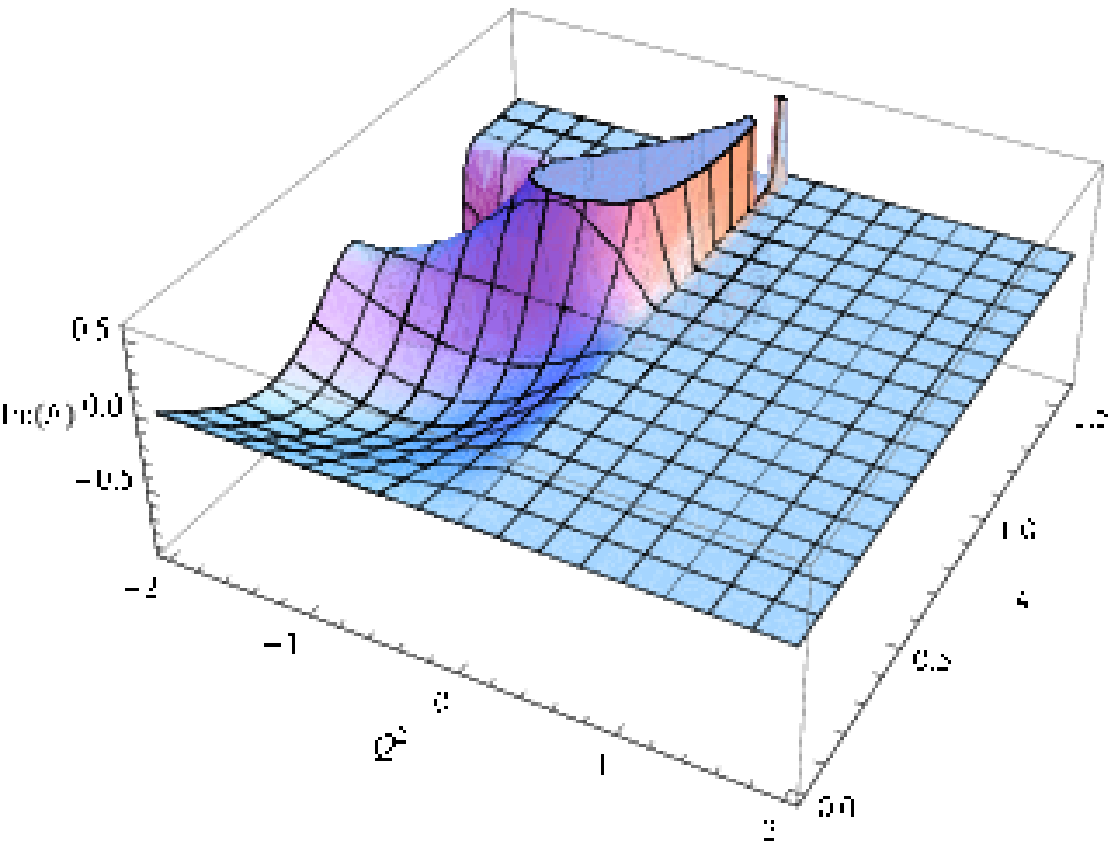}
\hspace{0.5cm}
\includegraphics[height=1.5in,width=1.5in]{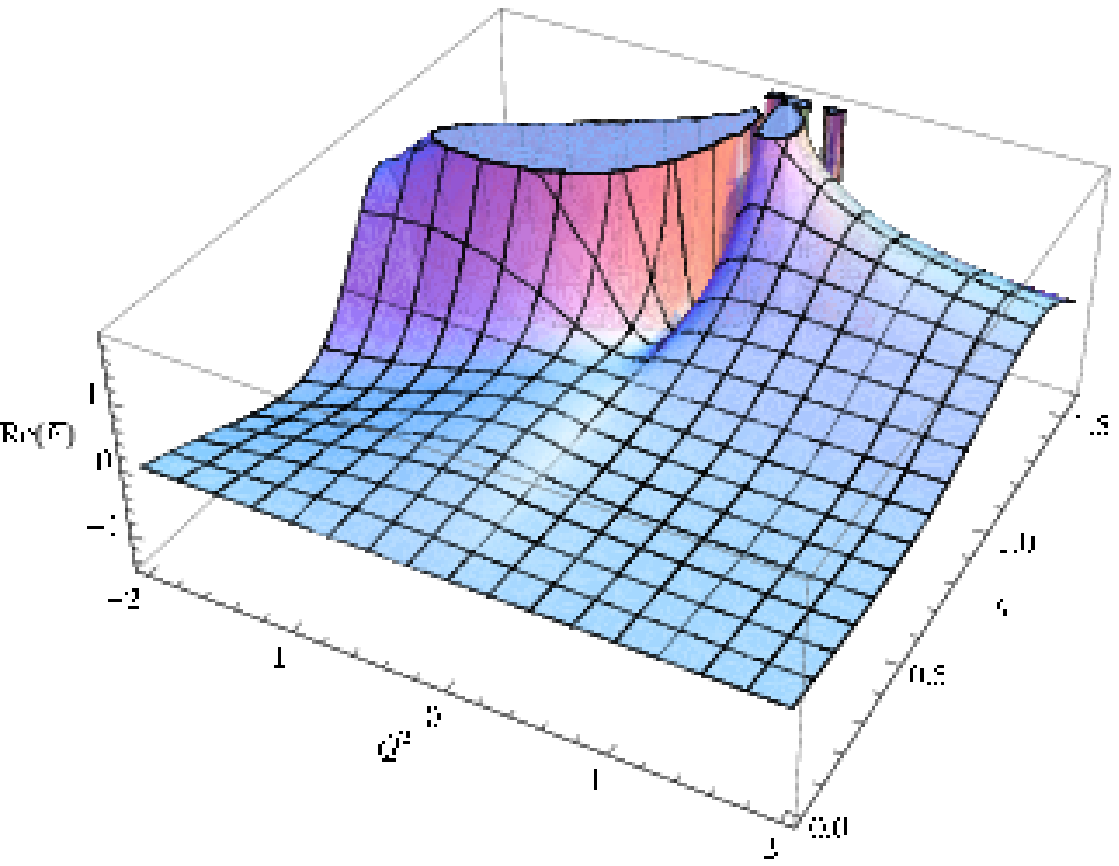}
\caption{Spin effect on the space- and time-like pion form factor respecting
conformal symmetry($m=0$) in the range of
$-2 {\rm GeV}^2\leq Q^2\leq 2 {\rm GeV}^2$.
Upper and lower panels represent the scalar($s=0$) and pseudoscalar($s=1/2$)
pion cases, respectively.}
\label{fig01}
\end{figure}
In Fig.~\ref{fig01}, we show the spin effect on the space- and time-like form
factor  respecting conformal symmetry($m=0$) in the range of
$-2 {\rm GeV}^2\leq Q^2\leq 2 {\rm GeV}^2$.
Upper and lower panels represent the scalar ($s=0$) and pseudoscalar ($s=1/2$)
pion cases, respectively.
For a fixed $Q^2$ value, $\theta$ approaches to $0$ and $\pi/2$ for the
low and high transverse momentum square ${\bf k}_\perp^2$, respectively.
Fig.~\ref{fig01} reveals that the high ${\bf k}_\perp^2$ region contributes
more to the form factor than the low ${\bf k}_\perp^2$ region.
This feature is more enhanced by the spin effect as one can see from the
comparison between the upper and lower panels of Fig.~\ref{fig01}.
Only after the $\theta$ integration, we are able to check the dispersion relation
between the real and imaginary parts of the form factor as we show
in Fig.~\ref{fig05}.

So far, we discussed the spin effect on the pion form factor in the
conformal limit. In the rest of this section, we discuss
the pion form factor given by Eq.~(\ref{soft_ff}) considering the
momentum-dependent dynamical quark mass $m(Q^2)$ used in our previous
work~\cite{CJ06}. Unlike the massless quark case, it is not so easy to analyze
the time-like region with Eq.~(\ref{soft_ff}) by doing analytic continuation
due to the functional form of $m(Q^2)$ taken.
\begin{figure}
\vspace{0.8cm}
\includegraphics[height=3in,width=3in]{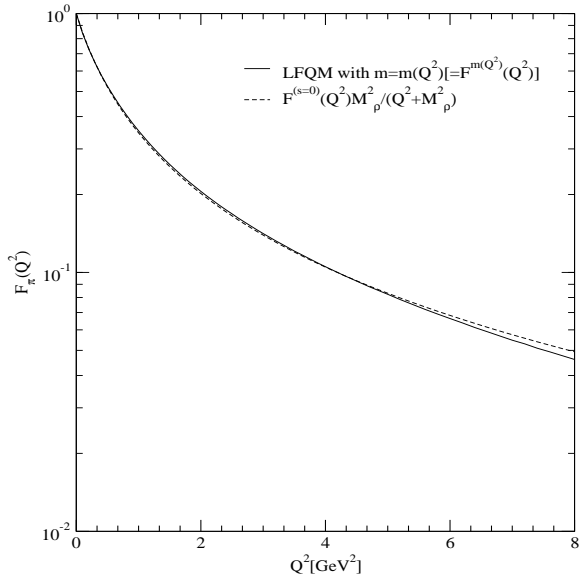}
\caption{Space-like pion form factor obtained from $F^{m(Q^2)}_\pi(Q^2)$[
Eq.~(\ref{soft_ff})] with the dynamical quark mass $m(Q^2)$ (solid line)
compared to the modified soft-wall AdS/QCD result~[Eq.~(\ref{AdS2})] with $\rho$ pole(dashed line).}
\label{fig02}
\end{figure}
Nevertheless, as one can see from Fig.~\ref{fig02}, we find that
Eq.~(\ref{soft_ff}) with $m=m(Q^2)$ is well approximated by the following
analytic form up to intermediate space-like momentum transfer region:
\be\label{AdS2}
F^{m(Q^2)}_{\pi}(Q^2)\simeq \frac{M^2_\rho}{Q^2 + {\cal M}^2_\rho}
F^{s=0}_\pi(Q^2),
\ee
where $M_\rho$ is the physical mass of the $\rho$ meson.
For convenience, we shall call the right-hand-side of Eq.~(\ref{AdS2}) as the
modified soft-wall AdS/QCD result with $\rho$ pole. We thus use the modified
soft-wall AdS/QCD result with $\rho$ pole as an approximate solution to the
time-like pion form factor $F^{m(q^2)}_\pi(q^2)$ with the dynamical quark mass.
We should note in the calculation of the time-like pion form factor from
Eq.~(\ref{AdS2}) that $M^2_\rho$ is replaced by
${\cal M}^2_\rho=M^2_\rho - iM_\rho\Gamma(q^2)$ in the denominator
where $M_\rho=776$ MeV and $\Gamma=120$ MeV to compare with the time-like pion
form factor data near the $\rho$ peak.

\section{Numerical results}
In this section, we compare our results for the space- and time-like form factor
with the experimental data. We also show that our direct calculation of the
time-like form factor obtained by the analytic continuation from the space-like
region is in excellent agreement with that obtained from the dispersion relation.

The model parameters used in our numerical calculations are the followings:
$\beta=0.173$ GeV for $F^{s=0}_\pi(Q^2)$, $\kappa=0.4$ GeV for
$F^{\rm AdS/QCD}_{(u.c.)}$~\cite{BT07}, $\beta=0.39$ GeV for
$F^{s=1/2}_\pi(Q^2)$ GeV and $F^{m(Q^2)}_\pi(Q^2)$(see~\cite{CJ06} for the details
of phenomenological form of the quark mass evolution), and $\kappa=3.2$ GeV
for the modified AdS/QCD result with $\rho$-pole. With these model parameters,
we are able to show the equivalences
$F^{s=0}_\pi(Q^2)\simeq F^{\rm AdS/QCD}_{(u.c.)}(Q^2)$ and
$F^{m(Q^2)}_\pi(Q^2)\simeq \frac{M^2_\rho}{ Q^2 + {\cal M}^2_\rho}
F^{s=0}_\pi(Q^2)$.
We obtain $F^{m(Q^2)}_\pi(Q^2)$ by systematically taking into account
the spin and quark mass evolution effects to $F^{s=0}_\pi(Q^2)$ and verify
the equivalence numerically
modifying the form factor of the soft-wall unconfined current ($F^{\rm
AdS/QCD}_{(u.c.)}(Q^2)$) with the empirical $\rho$-pole factor.
We should note that
the value of parameter $\kappa$ in the modified soft-wall AdS/QCD result
can be drastically different from that of $F^{\rm AdS/QCD}_{(u.c.)}(Q^2)$ due to
the modification by the empirical $\rho$-pole factor.
However, the physical observables are still comparable between the two models
with and without the modification.
For example, we obtain the pion decay constant $f_\pi=82.7$ MeV for
$F^{m(Q^2)}_\pi(Q^2)$ compared to 86.6 MeV for $F^{\rm AdS/QCD}_{(u.c.)}(Q^2)$
with $\kappa=0.4$ GeV.
\begin{figure}
\vspace{0.8cm}
\includegraphics[height=3in,width=3in]{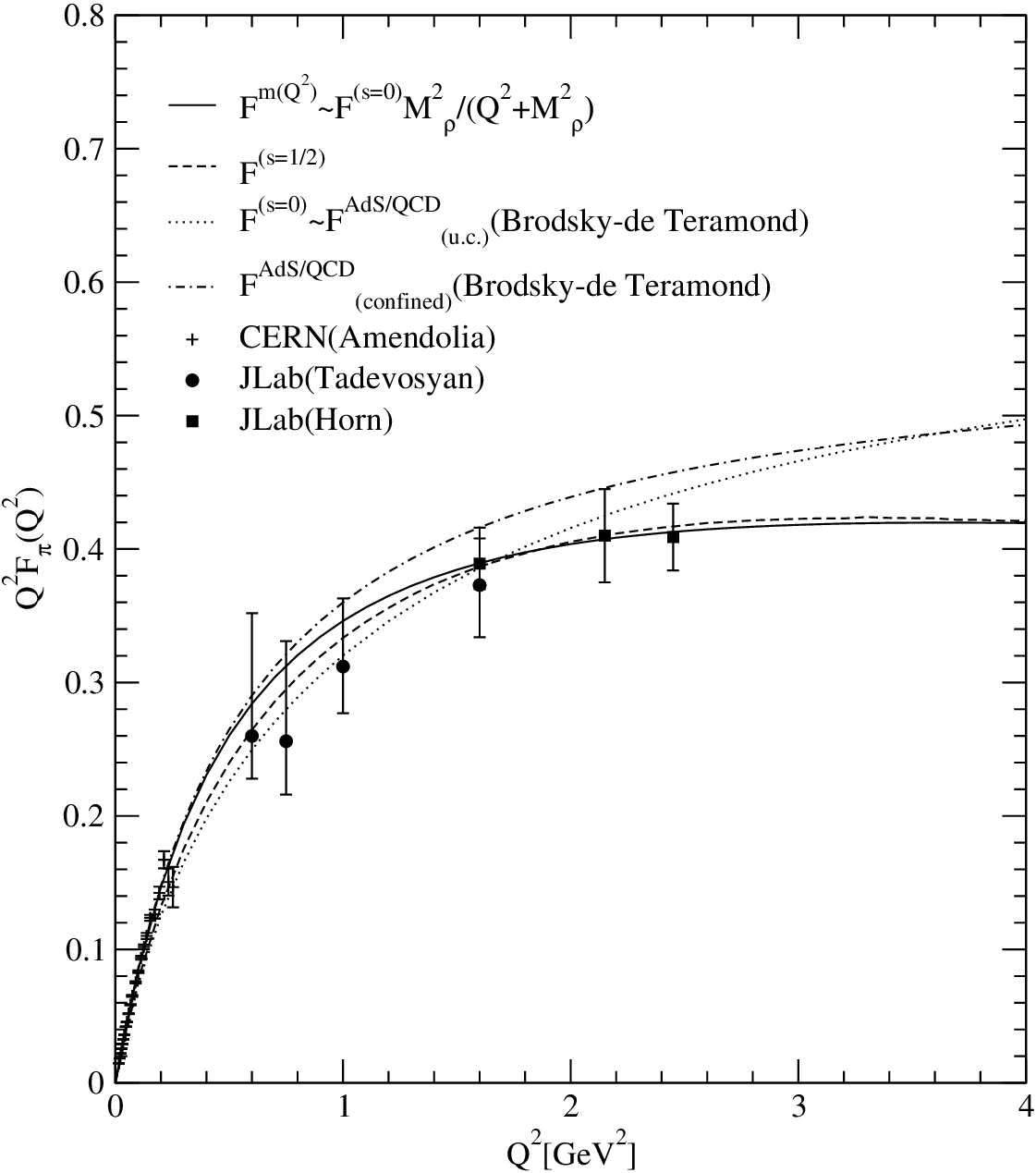}
\caption{Space-like behavior of $Q^2F_\pi(Q^2)$ for $0\leq Q^2\leq 4$ GeV$^2$
region.  Data are
taken from~\cite{Bebek,Amen,Jlab061,Jlab062,Jlab063}.}
\label{fig03}
\end{figure}
In Fig.~\ref{fig03}, we show the space-like behavior of $Q^2F_\pi(Q^2)$ for
$0\leq Q^2\leq 4$ GeV$^2$ region. The solid, dashed,  dotted, and dot-dashed
lines represent our LFQM result with the dynamical quark mass
$m(Q^2)$ in Eq.(\ref{soft_ff})(or
equivalently modified soft-wall AdS/QCD
model with $\rho$-pole in Eq.~(\ref{AdS2})),
the pseudoscalar pion form factor $F^{s=1/2}_\pi(Q^2)$ with $m=0$,
the scalar pion form factor $F^{s=0}_\pi(Q^2)$ with $m=0$(or eqivalently
$F^{\rm AdS/QCD}_{(u.c.)}$ with the unconfined current in Eq.~(\ref{AdS1})), and the
AdS/QCD result $F^{\rm AdS/QCD}_{(confined)}(Q^2)$ with the confined current
in Eq.~(\ref{AdS_c}),
respectively. Data are taken from~\cite{Bebek,Amen,Jlab061,Jlab062,Jlab063},
which includes the most recent results from JLAB~\cite{Jlab063}.
While our previous LFQM result~\cite{CJ99,CJ06} with
constituent constant quark mass $m=220$ MeV shows the gaussian
fall-off at high $Q^2$ region, all the lines in
Fig.~\ref{fig03} show the power-law behaviors at the available $Q^2$ scale.
Our $Q^2F^{m(Q^2)}_\pi(Q^2)$ with the dynamical quark mass(solid line) fits
the data not only for the very low $Q^2$ region but also for the intermediate
$Q^2$ region. Our $Q^2F^{s=1/2}_\pi(Q^2)$(dashed line) is the result obtained
by taking $m=0$ in $Q^2F^{m(Q^2)}_\pi(Q^2)$. So the little difference between
the two results shows the dynamical quark mass effect in the pion.
Similarly, the difference between the result of the AdS confined current
$Q^2F^{\rm AdS/QCD}_{(confined)}$(dot-dashed line) and that of the AdS unconfined
current $Q^2F^{\rm AdS/QCD}_{(u.c.)}(Q^2)\simeq Q^2F^{s=0}_\pi(Q^2)$(dotted line)
in low $Q^2$ region accounts for the effect of the dilaton field
discussed in~\cite{BT_Pi}.

\begin{figure}
\vspace{0.8cm}
\includegraphics[height=3in,width=3in]{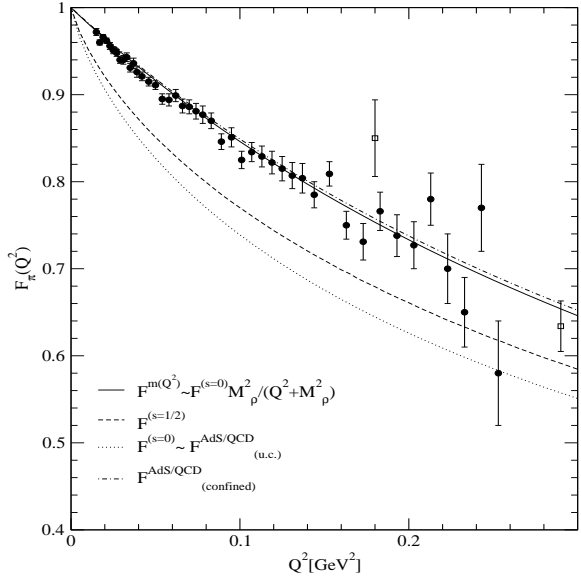}
\caption{Space-like behavior of $F_\pi(Q^2)$ for $0\leq Q^2\leq 0.3$ GeV$^2$
region.  Data are taken from~\cite{Bebek,Amen}.}
\label{fig04}
\end{figure}
In Fig.~\ref{fig04}, we show the space-like behavior of $F_\pi(Q^2)$
for low $0\leq Q^2\leq 0.3$ GeV$^2$ region. The same line codes are used as in
Fig.~\ref{fig03}. Our $F^{m(Q^2)}_\pi(Q^2)$(solid line) and
$F^{\rm AdS/QCD}_{(confined)}(Q^2)$(dot-dashed line)
for this low momentum transfer region are not only very close to each other
but also go through the data quite well.
However, both $F^{s=0}_\pi(Q^2)$(i.e. soft-wall AdS/QCD model with the unconfined
current) and $F^{s=1/2}_\pi(Q^2)$ with $m=0$ overestimate the pion charge radius.
From Figs.~\ref{fig03} and \ref{fig04}, we find that the low $Q^2$
behaviors between $F^{m(Q^2)}_\pi(Q^2)$ and $F^{\rm AdS/QCD}_{(confined)}(Q^2)$
are quite comparable to each other although the power-law behaviors at large
momentum transfer region are different, i.e.
$F^{m(Q^2)}_\pi(Q^2)\sim 1/Q^4$ vs. $F^{\rm AdS/QCD}_{(confined)}(Q^2)\sim 1/Q^2$.
This may indicate a correspondence between the dilaton effect
in AdS space and the spin and the mass evolution effects of constituent quark
and anti-quark inside the pion.

\begin{figure}
\vspace{0.8cm}
\includegraphics[height=3in,width=3in]{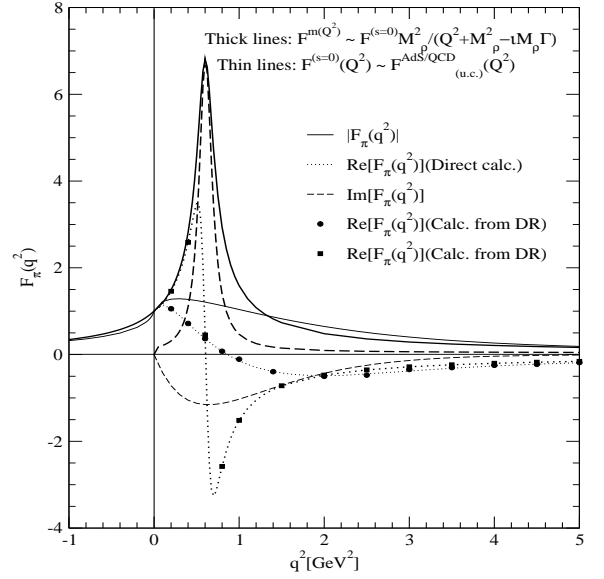}
\caption{Space- and time-like pion form factor obtained from
$F^{m(Q^2)}_\pi(Q^2)\simeq\frac{M^2_\rho}{Q^2+{\cal M}^2_\rho}
F^{s=0}_\pi(Q^2)$ (thick lines) and
$F^{s=0}_\pi(Q^2)\simeq F^{\rm AdS/QCD}_{(u.c)}(Q^2)$ GeV(thin lines)
compared with the results obtained from the dispersion relation(black data).}
\label{fig05}
\end{figure}
In Fig.~\ref{fig05}, we show the pion form factor for both space- and time-like
region obtained from
$F^{m(Q^2)}_\pi(Q^2)\simeq\frac{M^2_\rho}{Q^2+{\cal M}^2_\rho}
F^{s=0}_\pi(Q^2)$ (thick lines) and
$F^{s=0}_\pi(Q^2)\simeq F^{\rm AdS/QCD}_{(u.c.)}(Q^2)$(thin lines),
respectively. The solid, dotted, and dashed lines
represent $|F_\pi(q^2)|$, Re[$F_\pi(q^2)$], and Im[$F_\pi(q^2)$], respectively. The
black circles and squares are the results of Re[$F^{s=0}_\pi(q^2)$] and
Re[$F^{m(q^2)}_\pi(q^2)$] obtained from the dispersion relation given by
Eq.~(\ref{Re_F}).
Fig.~\ref{fig05} shows that our direct calculations are in an excellent agreement
with the solutions of the dispersion relation. Although
$F^{s=0}_\pi(Q^2)$ in time-like region produces a $\rho$ meson-type peak near
$q^2\sim M^2_\rho$ as in the case of our previous scalar field theory
model~\cite{CJ_Dis}, it does not yield
all the features of the vector meson dominance phenomena as
$F^{m(Q^2)}_\pi(Q^2)\simeq\frac{M^2_\rho}{Q^2+{\cal M}^2_\rho}F^{s=0}_\pi(Q^2)$
does. This indicates that the spin and mass
evolution effects are crucial in generating the
more realistic features of the vector meson dominance phenomena.

\begin{figure}
\vspace{0.8cm}
\includegraphics[height=3in,width=3in]{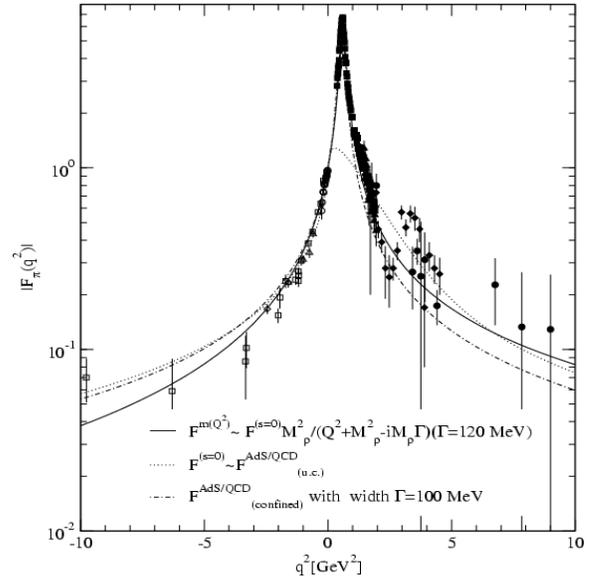}
\caption{Space- and time-like pion form factor obtained from
$F^{m(Q^2)}_\pi(Q^2)\simeq\frac{M^2_\rho}{Q^2+{\cal M}^2_\rho}F^{s=0}_\pi(Q^2)$
(solid line),
$F^{s=0}_\pi(Q^2)\simeq F^{\rm AdS/QCD}_{(u.c.)}(Q^2)$ GeV(dotted line),
and
$F^{\rm AdS/QCD}_{(confined)}(Q^2)$(dot-dashed line),
respectively. The data are taken
from~\cite{Bebek,Amen,Jlab061,Jlab062,Jlab063,OL1,OL2,DM2,BCF}. }
\label{fig06}
\end{figure}

In Fig.~\ref{fig06}, we show the pion form factor for both space- and time-like
region obtained from
$F^{m(Q^2)}_\pi(Q^2)\simeq\frac{M^2_\rho}{Q^2+{\cal M}^2_\rho}F^{s=0}_\pi(Q^2)$
(solid line),
$F^{s=0}_\pi(Q^2)\simeq F^{\rm AdS/QCD}_{(u.c.)}(Q^2)$ GeV(dotted line),
and
$F^{\rm AdS/QCD}_{(confined)}(Q^2)$~\cite{BT_Pi}(dot-dashed line),
respectively. The data in space- and time-like regions are taken
from~\cite{Bebek,Amen,Jlab061,Jlab062,Jlab063} and from~\cite{OL1,OL2,DM2,BCF},
respectively.  As one can see from Fig.~\ref{fig06},
our $F^{m(Q^2)}_\pi(Q^2)$ and $F^{\rm AdS/QCD}_{(confined)}(Q^2)$~\cite{BT_Pi}
exhibit more realistic $\rho$ meson-type peak than $F^{s=0}_\pi(Q^2)$.
Our LFQM analysis indicates that the difference between
$F^{m(Q^2)}_\pi(Q^2)$ and $F^{s=0}_\pi(Q^2)$
is due to the spin(i.e. Melosh factor) and dynamical mass evolution effects of the
constituent quark and anti-quark inside the pion. Our result may also be compared to
the difference between $F^{\rm AdS/QCD}_{(confined)}(Q^2)$ and
$F^{\rm AdS/QCD}_{(u.c.)}(Q^2)$, which is due to the effect of the dilaton field
in AdS space~\cite{BT_Pi}.

\section{Summary and Conclusion}
We discussed a constraint of conformal symmetry in the analysis of the pion
form factor. Working in the framework of the LFQM that takes into
account a momentum-dependent dynamical quark mass~\cite{CJ06}, we
extended our previous analysis of the space-like pion form factor to the
time-like region and compared with the result obtained from the AdS/QCD
model~\cite{BT_Pi}. We showed explicitly that the time-like form factor obtained by the
analytic continuation of the space-like form factor correctly satisfies the dispersion
relation. Our comparative analysis between the scalar quark case ($F^{s=0}_\pi(Q^2)$)
and the spin $1/2$ dynamical quark mass case ($F^{m(Q^2)}_\pi(Q^2)$) indicates that
the quark spin and dynamical mass effects are crucial to yield the realistic features of
the vector meson dominance phenomena.

\acknowledgments
We thank Stan Brodsky and Guy de Teramond for the useful discussion
during the BARYONS07 International Conference and the subsequent correspondence.
This work was supported by a grant from the U.S. Department of
Energy(No. DE-FG02-96ER40947).  H.-M. Choi was supported in part by Korea
Research Foundation under the contract KRF-2005-070-C00039 and in part by Kyungpook National
University Reserch Fund, 2007. The National
Energy Research Scientific Center is also acknowledged for the grant of
supercomputing time.


\begin{thebibliography}{99}
\bibitem{Mal99} J. M. Maldacena,  Adv. Theor. Math. Phys. {\bf 2}, 231
(1998)[Int. J. Theor. Phys. {\bf 38}, 1113 (1999)];
S. S. Gubser, I. R. Klebanov, and A. M. Polyakov,
\Journal{\PLB}{428}{105}{1998}; E. Witten, Adv. Theor.
Math. Phys. {\bf 2}, 253 (1998).
\bibitem{Mass} J. Erlich, E. Katz, D. T. Son, and
M. A. Stephanov, \Journal{\PRL}{95}{261602}{2005}; H. Boschi-Filho and
N. R. F. Braga, JHEP {\bf 0305}, 009 (2003); N. Evans and A. Tedder,
\Journal{\PLB}{642}{546}{2006}; D. K. Hong, T. Inami and H. U. Yee,
\Journal{\PLB}{646}{165}{2007}; P. Colangelo, F. De Fazio, F. Jugeau and
S. Nicotri, \Journal{\PLB}{652}{73}{2007}; H. Forkel, M. Beyer and T.
Frederico, JHEP {\bf 0707}, 077 (2007).
\bibitem{PS02} J. Polchinski and M. J. Strassler,
\Journal{\PRL}{88}{031601}{2002}; JHEP {\bf 0305}, 012 (2003).
\bibitem{BT04} S. J. Brodsky and G. F. de Teramond,
\Journal{\PLB}{582}{211}{2004}; G. F. de Teramond and
S. J. Brodsky, \Journal{\PRL}{94}{201601}{2005}.
\bibitem{BL} G. P. Lepage and S. J. Brodsky, \Journal{\PRD}{22}{2157}{1980}.
\bibitem{BT06} S. J. Brodsky and Guy F. de Teramond,
\Journal{\PRL}{96}{201601}{2006}.
\bibitem{Rad06} A. V. Radyushkin, \Journal{\PLB}{642}{459}{2006}.
\bibitem{GuyStan} G. F. de Teramond, Talk entitled
``Mapping AdS/CFT Results forHolographic QCD to the Light Front"
presented in the Workshop LC2006 (Light-Cone QCD and Nonperturbative
Hadron Physics), Minneapolis, May 15-19, 2006; S. J. Brodsky,
hep-ph/0608005 and hep-ph/0610115.
\bibitem{BT07} S. J. Brodsky and Guy F. de Teramond, arXiv:hep-ph/0702205.
\bibitem{BT_Pi} S. J. Brodsky and Guy F. de Teramond,
\Journal{\PRD}{77}{056007}{2008}(arXiv:0707.3859[hep-ph]).
\bibitem{GR07} H. R. Grigoryan and A. V. Radyushkin,
\Journal{\PLB}{650}{421}{2007}; \Journal{\PRD}{76}{095007}{2007}.
\bibitem{KL} H. J. Kwee and R. F. Lebed, arXiv:0708.4054[hep-ph];
arXiv:0712.1811[hep-ph].
\bibitem{KK} A. Karch, E. Katz, D. T. Son, and M. A. Stephanov,
\Journal{\PRD}{74}{015005}{2006}.
\bibitem{CJ06} H. M. Choi and C. R. Ji, \Journal{\PRD}{74}{093010}{2006}.
\bibitem{CJ99} H.-M. Choi and C.-R. Ji, \Journal{\PRD}{59}{074015}{1999}.
\bibitem{Mel} H. J. Melosh, \Journal{\PRD}{9}{1095}{1974}.
\bibitem{CJ_Jacob} H.-M. Choi and C.-R. Ji, \Journal{\PRD}{56}{6010}{1997}.
\bibitem{Wil}
J.I. Skullerud and A.G. Williams, \Journal{\PRD}{63}{054508}{2001};
J.B. Zhang, P.O. Bowman, D.B. Leinweber, A.G. Williams and F.D.R. Bonnet,
\Journal{\PRD}{70}{034505}{2004}.
\bibitem{DS1} P. Maris and C.D. Roberts, \Journal{\PRC}{58}{3659}{1998}.
\bibitem{DS2} P. Maris and P.C. Tandy, \Journal{\PRC}{62}{055204}{2000}.
\bibitem{Dor} A. E. Dorokhov and L. Tomio, \Journal{\PRD}{62}{014016}{2000};
A. E. Dorokhov, \Journal{\EPJC}{32}{79}{2003}.
\bibitem{GP} T. Gousset and B. Pire, \Journal{\PRD}{51}{15}{1995}.
\bibitem{Bebek} C. J. Bebek {\em et al.}, \Journal{\PRD}{17}{1693}{1978}.
\bibitem{Amen} S.R. Amendolia {\em et al.}, \Journal{\NPB}{277}{168}{1986}.
\bibitem{Jlab061} T. Horn {\em et al.}, \Journal{\PRL}{97}{192001}{2006}.
\bibitem{Jlab062} V. Tadevosyan {\em et al.}, nucl-ex/0607007.
\bibitem{Jlab063} T. Horn {\em et al.}, arXiv:0707.1794[nucl-ex].
\bibitem{CJ_Dis} H.-M. Choi and C.-R. Ji, \Journal{\NPA}{679}{735}{2001}.
\bibitem{OL1} OLYA Collaboration, A. D. Bukin {\em et al.},
\Journal{\PLB}{73}{226}{1978}.
\bibitem{OL2} OLYA and CMD Collaborations, L. M. Barkov {\em et al.},
\Journal{\NPB}{256}{365}{1985}.
\bibitem{DM2} DM2 Collaboration, D. Bisello {\em et al.},
\Journal{\PLB}{220}{321}{1989}.
\bibitem{BCF} BCF Collaboration, D. Bollini {\em et al.},
Lett. Nuovo Cim. {\bf 14}, 418 (1975).
\end{thebibliography}
\end{document}